\documentclass[%
 reprint,
 amsmath,amssymb,
 aps,
]{revtex4-2}

\usepackage{graphicx}
\usepackage{amsmath}
\usepackage{nicefrac}
\usepackage[hidelinks]{hyperref}
\usepackage{braket}
\usepackage{amssymb}
\usepackage{siunitx}
\DeclareSIUnit\gauss{G}
\usepackage{xspace}
\usepackage{mathtools}
\usepackage{tikz}

\newcommand{\swap}{\ifmmode\text{\textsc{swap}}\else\textsc{swap}\fi\xspace}
\newcommand{\swapsq}{\ifmmode\text{\textsc{swap}}^2\else$\text{\textsc{swap}}^2$\fi\xspace}
\newcommand{\swapd}{\ifmmode\sqrt{\text{\sc swap}}^{\dagger}\else$\sqrt{\text{\sc swap}}^{\dagger}$\fi\xspace}
\newcommand{\swapa}{\ifmmode(\text{\sc swap})^{\alpha}\else$(\text{\sc swap})^{\alpha}$\fi\xspace}
\newcommand{\sswap}{\ifmmode\sqrt{\text{\sc swap}}\else$\sqrt{\text{\sc swap}}$\fi\xspace}
\DeclareRobustCommand{\singlet}{\ensuremath{\mathrel{\,\tikz[baseline=-0.5ex]{\draw[thick] (0,0) -- (2.5ex,0);\draw[thick, fill=white] (0,0) circle (0.5ex);\draw[thick, fill=white] (2.5ex,0) circle (0.5ex);}\,}}}

\begin{document}

\title{Digital programming of spin correlations in a fermionic lattice quantum simulator}

\author{Yann Kiefer}
\thanks{These authors contributed equally}
\author{Lars Fischer}
\thanks{These authors contributed equally}
\author{Zijie Zhu}
\thanks{These authors contributed equally. Present address: Department of Physics, Fudan University, Shanghai, China}
\author{Konrad Viebahn}
\email[]{viebahnk@phys.ethz.ch}
\author{Tilman Esslinger}
\affiliation{Institute for Quantum Electronics \& Quantum Center, ETH Zurich, 8093 Zurich, Switzerland}

\date{\today}

\begin{abstract}
Analog quantum simulation provides a highly controlled platform to study diverse quantum many-body phenomena. However, current methods for state initialisation are limited to thermal ensembles or uncorrelated product states. Here we present a hybrid approach that complements analog preparation with a digital quantum-gate protocol. This approach enables the engineering of target states with specific, long-range spin-correlations from the same initial resource state. By applying collisional gates to adiabatically prepared and filtered four-fermion singlet chains, we program diverse spin-correlation patterns, including that of a Heisenberg chain. We measure the spin correlations using a sequence of quantum gates followed by singlet-pair measurements. Our method paves the way to the targeted preparation of strongly correlated states of matter. 
\end{abstract}

\maketitle

\textit{Introduction}---The deterministic preparation of strongly correlated quantum states is a key challenge of quantum simulation ~\cite{georgescu_quantum_2014,altman_quantum_2021, daley_practical_2022}. Reaching these states is of particular interest in fermionic systems, where the combination of Pauli exclusion, strong interactions, and entanglement gives rise to phenomena such as unconventional superconductivity, frustrated magnetism, and topological order. State preparation in a quantum simulator typically relies on one of two complementary approaches. The first is the analog approach, where correlations develop through Hamiltonian evolution either from a thermal ensemble \cite{greif_short-range_2013,hart_observation_2015,boll_spin-_2016,cheuk_observation_2016,mazurenko_cold-atom_2017,brown_spin-imbalance_2017,gorg_enhancement_2018,gall_competing_2021,christakis_probing_2023,shao_antiferromagnetic_2024,xu_neutral-atom_2025} or from an uncorrelated product state \cite{fukuhara_quantum_2013,preiss_strongly_2015,islam_measuring_2015,labuhn_tunable_2016,bernien_probing_2017,brydges_probing_2019,bayha_observing_2020,dehollain_nagaoka_2020,leonard_realization_2023,lunt_realization_2024,sun_experimental_2026}. The second is the digital approach, where correlations are built using gate sequences \cite{houck_-chip_2012,zhang_scalable_2023, nigmatullin_experimental_2025}. In the analog method, the physics of the target Hamiltonian naturally emerges through adiabatic evolution, which is, however, limited by finite temperature and closing energy gaps. While digital methods offer programmable control, they demand a high level of connectivity and coherence for the simulation of fermionic systems, which current platforms lack. Although hybrid analog-digital strategies have recently been implemented on different platforms \cite{impertro_local_2024,evered_probing_2025,andersen_thermalization_2025,senoo_high-fidelity_2025} to address these limitations, preparing strongly correlated fermionic states in a deterministic and programmable manner remains an open challenge. The ability to deterministically prepare such states promises vital insights into condensed matter~\cite{bohrdt_exploration_2021,qin_hubbard_2022,arovas_hubbard_2022}, quantum chemistry \cite{cao_quantum_2019} and gauge theories~\cite{halimeh_cold-atom_2025}.

\begin{figure*}
    \includegraphics[width=0.95\textwidth]{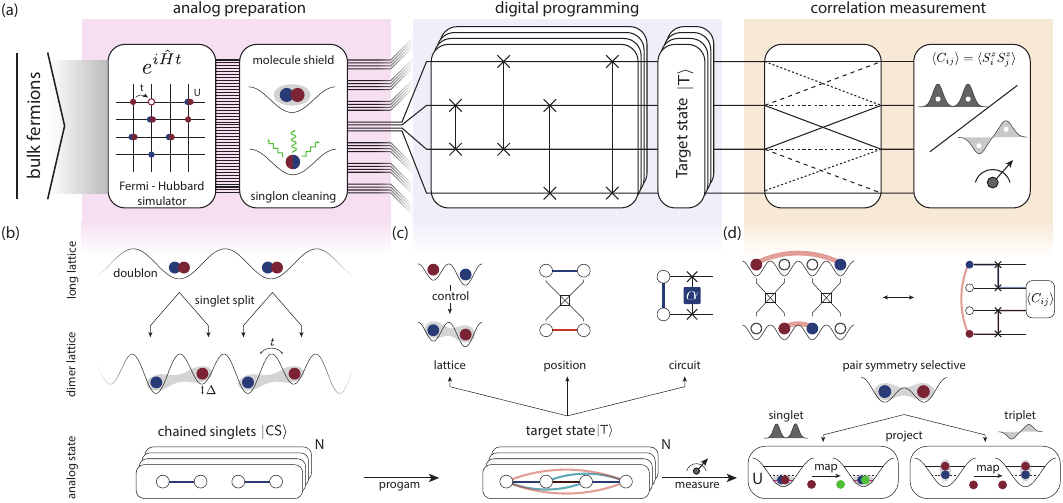}
         \caption{\textbf{Programming spin correlations in a hybrid Fermi-Hubbard simulator.} (a) Schematic illustration of the analog-digital protocol to program four-particle quantum correlations. Concatenation of analog and digital methods enables configurable entanglement structures approaching a specific target state $\ket{\mathsf{T}}$. The configured correlations are measured with a protocol combining atom-rearrangement and a symmetry-selective detection procedure. (b) Analog state preparation is enabled by bulk evolution of a degenerate Fermi gas in a tunable optical superlattice. Attractive interactions during loading the long lattice achieve a large doublon occupation. Remaining singlons are removed via resonant light and molecular shielding. The doublons are split into atomic singlets $\singlet = (\ket{\uparrow\downarrow} - \ket{\downarrow\uparrow})/\sqrt{2}$ ramping up a short wavelength lattice. Singlon removal in the long lattice ensures isolation of $N$ chained singlets $\ket{\mathsf{CS}}=(\singlet \singlet)$. (c) Chained singlets are configured via tunable digital gates based on atomic collisions within the lattice. Control of the energy offset $\Delta$ and tunnelling $t$ ensures motional confinement during the programming sequence and the deterministic configuration of a target state $\ket{\mathsf{T}}$. (d) Local measurement of the spin pair-correlators  $\langle C_{ij}\rangle=\langle S_i^zS_j^z\rangle 4/\hbar^2$. Rearrangement of $\ket{\mathsf{T}}$ prior to measurement is achieved via topological pumping in combination with high-fidelity collisional gates. After rearrangement, the spin correlators are measured with a wavefunction symmetry-selective procedure.}
    \label{fig:1}
\end{figure*}

In this work, we build strongly correlated quantum states by combining analog methods of bulk Hamiltonian evolution with programmable digital gate sequences based on entangling atomic collisions \cite{jaksch_entanglement_1999,brennen_quantum_1999,mandel_controlled_2003,anderlini_controlled_2007,kaufman_entangling_2015,yang_cooling_2020,zhu_splitting_2025,bojovic_high-fidelity_2026} and geometric gates \cite{kiefer_protected_2026}. Using analog state preparation as a first step, we isolate four-fermion chains as fundamental building blocks and apply programmable gate-based circuits to encode their correlation structure.
With this method we are able to prepare quantum correlations that are not native to the underlying Hamiltonian. This includes states with non-local configurations, which we can verify using coherent rearrangement and local two-body readout.  

Our implementation begins by loading $9.53(30) \times 10^{4}$ atoms of a two-component degenerate Fermi gas ($^{40}\text{K}$, states $\ket{\uparrow} = \ket{F=9/2, m_F=-7/2}$ and $\ket{\downarrow} = \ket{F=9/2, m_F=-9/2}$) into a three-dimensional dynamical optical superlattice \cite{walter_quantization_2023}, that forms an ensemble of independent one-dimensional tubes (Figure~\ref{fig:1}(a)). We initialize the resource state by applying attractive interactions between atoms in $\ket{\uparrow}$ and $\ket{\downarrow}$ states during loading into the long-period lattice, such that doublons form (i.e. atoms occupying the same site with total spin $S=0$). Next, we purify the doublon filling by removing atoms on singly occupied lattice sites (singlons). For this, we first turn doublons into Feshbach molecules, which shifts their resonance with respect to the resonances of the singlons. Resonant light at the singlon transition frequency then selectively addresses and removes the singlon defects from the system (Methods~\ref{SMsec:singletDist}, see also refs.~\cite{xu_formation_2003,krauser_coherent_2012}). This procedure ensures that the doublons are either adjacent to another doublon or to a vacancy. Following this purification, we adiabatically split the doublons into atomic singlets residing on double-wells $\singlet = (\ket{\uparrow\downarrow} - \ket{\downarrow\uparrow})/\sqrt{2}$ by ramping up a short-wavelength lattice in the presence of repulsive interactions (Figure ~\ref{fig:1}(b)) \cite{lubasch_adiabatic_2011, zhu_splitting_2025,xu_neutral-atom_2025}.
The ket $\ket{\uparrow\downarrow}$ denotes a state in which $\ket{\uparrow}$ occupies the left site and $\ket{\downarrow}$ occupies the right site of a two-site unit cell.
In summary, the preparation process yields a lattice in which $82.2(1)\%$ of the atoms are bound in singlet pairs, with more than $12(2)\%$ of the system forming four-particle chains (Methods \ref{SMsec:initialState}). These `chained singlets' serve as the primary resource states $\ket{\mathsf{CS}} = (\singlet \singlet)$ for our programmable architecture (Figure~\ref{fig:1}(b)).

We realise the digital programming sequence with a combination of spatial reconfiguration of the four fermion states using geometric and collisional entangling gates. Spatial reconfiguration of the system is based on bi-directional, state-independent atom shuttling using topological pumping \cite{zhu_splitting_2025,citro_thouless_2023}. This mechanism enables discrete and coherent transport of atoms to distinct lattice sites without destroying existing entanglement (Methods \ref{SMsec:latPot}).
Controlled atomic collisions implement tunable partial \swap gates (Figure \ref{fig:1}(c)), enabling the deterministic construction of specific correlation structures through entangling gate operations. Their dynamics is governed by a time-dependent Fermi-Hubbard Hamiltonian, parametrized by the dimer tunnelling $t$, the energy site offset $\Delta$, and the on-site Hubbard interaction $U$.
The phase acquired during one gate opration has a geometric of $\pi$ and an additional tunable dynamic phase contribution, both of which result from a dynamical ramp of $\Delta$ and $t$.

In the basis $\{\ket{\uparrow\uparrow}, \ket{\uparrow\downarrow},\ket{\downarrow\uparrow},\ket{\downarrow\downarrow}\}$, the combined operation is described by the unitary
\begin{align}
    \hat{U}({\alpha}) = \begin{pmatrix}
        1&0&0&0\\
        0&(1+e^{i\pi\alpha})/2 &(1-e^{i\pi\alpha})/2 &0\\
        0&(1-e^{i\pi\alpha})/2 &(1+e^{i\pi\alpha})/2 &0\\
        0&0&0&1
    \end{pmatrix},
\end{align}
where the total gate angle $\alpha = 1 + \frac{1}{\pi}\int_{\tau} J_{\text{ex}}(t)\, dt$ is continuously tunable via the Hubbard $U$ through a Feshbach resonance. Crucially, we operate in the direct-exchange regime $|U| \lesssim t$, where the dynamical phase exhibits a first-order dependence on $U$ \cite{kiefer_protected_2026}, in contrast to the superexchange regime ($U\gg t$) where $J_{ex}\propto t^2$ and the gate fidelity becomes sensitive to fluctuations in $t$. 
The case of a \swap gate is realized at $\alpha=1$ which corresponds to the non-interacting limit at $U=0$ and the gate phase is determined entirely by the geometric phase \cite{kiefer_protected_2026}. 
We program a deterministic digital circuit that acts globally on the system; however, because the chained- and isolated singlets are spatially decoupled by vacant sites, their quantum evolutions remain independent and do not interfere with each other (Methods \ref{SMsec:signalIsolation}). By composing these shuttling sequences with tunable collisional interactions, we demonstrate the ability to deterministically configure the many-body state, transforming initial local correlations into complex, non-local entanglement structures (Figure \ref{fig:1}(c)). 
\begin{figure*}
    \includegraphics[width=1\textwidth]{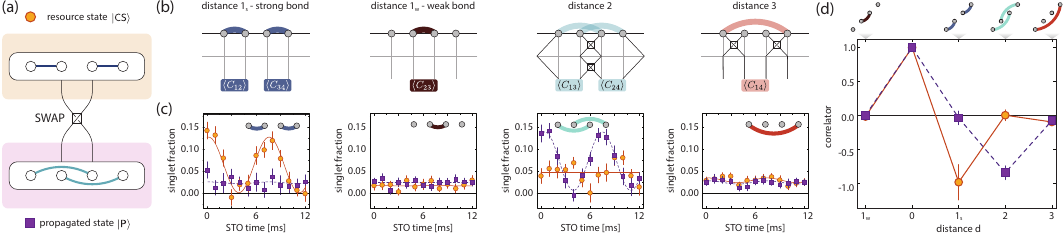}
         \caption{\textbf{Measurement of non-local quantum correlations} (a) Measurement of long-distance spin pair correlators illustrated on two distinct scenarios. The analog resource state $\ket{\mathsf{CS}}$ is compared to the case where the inner two atoms are swapped without changing the internal spin-configuration. This gives rise to the propagated singlet state $\ket{\mathsf{P}}$ in which the antiferromagnetic correlation has propagated to next-nearest neighbours. (b) Measurement of the spin-pair correlator $\langle C_{ij}\rangle$ schematics. The correlators are extracted in individual measurement sequences. For $d=1_{s,w}$ the appropriate superlattice bond is chosen by controlling the dynamical lattice ($s$ = strong, $w$ = weak). Correlators for distance $d>1$ require the rearrangement of atoms within the four-fermion chain as the projection of the spatial parity of the wavefunction requires atoms to be in the same lattice site. The black lines denote the trajectories of the atoms and crossing lines indicate a collisional gate. (c) Coherent singlet triplet oscillations of the resource- $\ket{\mathsf{CS}}$ (orange points) and propagated state $\ket{\mathsf{P}}$ (purple squares) after background subtraction. The propagated state is prepared by application a \swap gate between the inner two atoms realized with the unitary $\hat{U}(\pi)_{23}$. The correlator values are obtained by the fitted contrast of the oscillations (lines). The error bars denote the standard error for i=10-15 experimental realisations. (c) Extracted correlator values as a function of the distance for the resource (orange) and propagated (purple) bi-singlet state.}
    \label{fig:2}
\end{figure*}

To measure the spin pair-correlator, we employ a readout circuit that first uses topological pumping for atom-position reconfiguration followed by two-particle symmetry-selective measurement (Figure \ref{fig:1}(d)). Each pair-correlator is measured individually such that the two atoms of interest are brought to neighbouring sites. The two lattice sites are then merged, at which point fermionic statistics and the symmetry of the spatial wave function dictate the accessible motional states. Atoms in a spin-singlet configuration co-occupy the same energy level (spatially symmetric wavefunction), whereas atoms in a spin-triplet configuration must populate different bands. In the singlet case, the resulting on-site interaction shift provides a direct handle on the spin wavefunction symmetry. We exploit this with a conditional spin-flip operation that acts exclusively on the singlet subspace enabling the measurement of the singlet fraction \cite{greif_short-range_2013}.

\textit{Non-local spin correlator measurement -} One of the central components of our scheme is the capability to readout non-local quantum spin correlations $\langle C_{ij}\rangle = \frac{4}{\hbar^2}\langle S_i^zS_j^z\rangle$, where $\{ i,j\}$ denote the position of the atoms in each chain. To characterize the state, we implement a configurable readout circuit capable of measuring the spin-correlators $\langle C_{ij} \rangle$ of four-fermion states. We demonstrate the measurement protocol by comparing two distinct initial states prepared in the experiment. First, we measure the correlations of the chained-singlet state $\ket{\mathsf{CS}}=(\singlet_{12} \otimes \singlet_{34})$ initialized via adiabatic evolution. The second case is a propagated state of two singlets $\ket{\mathsf{P}} = (\singlet_{13} \otimes \singlet_{24})$, which is obtained by applying a non-entangling \swap gate between the central atoms (2 and 3) (Figure~\ref{fig:2}(a)).

While nearest-neighbor correlations ($d=\lvert i-j\rvert = 1$) are accessible by selecting the appropriate superlattice bond prior to measurement, long-distance correlators ($d \ge 2$) require the deterministic rearrangement of atoms to map non-local correlations onto adjacent sites (Figure \ref{fig:2}(b)). For next-nearest neighbor correlations ($d=2$), we exchange the positions of the inner two atoms (sites 2 and 3) while keeping the outer atoms stationary. This is realized by a sequence of two gates ($\alpha_1+\alpha_2=1$) separated by a reversal of the pump direction illustrated by the black lines, effectively realising a \swap gate between atoms located on sites 2 and 3. Distinct from schemes relying on the symmetric tunnelling in the lattice~\cite{impertro_local_2024,gkritsis_simulating_2025,roth_constructing_2026}, this approach prevents the diffusion of atoms located at the edges of the chain. Similarly, $d=3$ correlations are accessed via a high-fidelity geometric \swap gate connecting lattice sites 1 and 4. 

The measurement of the spin correlator is performed via coherent singlet-triplet oscillations (STOs) \cite{petta_coherent_2005,trotzky_controlling_2010,greif_short-range_2013}. For given pairs of atoms, we induce oscillations between the singlet $\ket{s}$ and triplet $\ket{t_0}=(\ket{\uparrow\downarrow}+\ket{\downarrow\uparrow})/\sqrt{2}$ states by applying a magnetic gradient $\Delta B$ for variable times $\tau$ (Figure~\ref{fig:2}(c)). Assuming SU(2) symmetry allows the extraction of the correlator value according to the contrast of this oscillation $\langle C_{ij}\rangle = (\mathcal{S}-\mathcal{T})/(3\mathcal{T}+\mathcal{S})$, where $\mathcal{S}$ and $\mathcal{T}$ denote the singlet populations at $\tau=0$ and $\tau=T/2$, respectively \cite{auerbach_interacting_1994}. Global readout necessitates a background subtraction of isolated singlets to obtain the signal from the four-body chains (see Methods \ref{SMsec:signalIsolation}). The measured correlations (Figure~\ref{fig:2}(d)) show that the singlet signal has shifted to the $\langle C_{13}\rangle$ and $\langle C_{24}\rangle$ sectors indicating successful propagation of the correlations. 

\begin{figure}
    \includegraphics[width=0.5\textwidth]{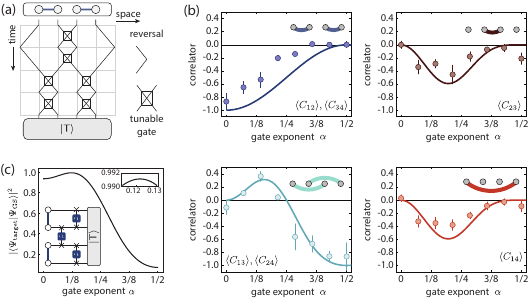}
         \caption{\textbf{Targeted preparation spin-correlations in fermionic singlet chains.} (a) schematic representation of the circuit for configuring a target state $\ket{\mathsf{T}}$ with non-local correlations starting form the chained singlet initial state $(\singlet \singlet)$. The analog resource is programmed with a total of four digital gate-operations characterised by their gate exponent $\alpha$. In the circuit, black lines denote trajectories of the atoms subject to topological pumping. Intersecting lines realize a collisional gate between atoms. The direction of the pump is reversed once after the first gate and again before the last gate to spatially collimate the four-fermion state. (b) Measured correlator values for $d=1_s$, $d=1_w$, $d=2$, and $d=3$ of $\ket{\mathsf{T}}$ resulting from the four-gate quantum circuit as a function of the gate exponent $\alpha$. The solid lines denote theory calculations in absence of any free parameters. The values are obtained from i=10-15 experimental realizations and the error bars denote the standard error. (c) Theory overlap of the target state $\ket{\mathsf{T}}$ with the ground state of the four particle state $\ket{\mathsf{GS}}$ as a function of the gate exponent $\alpha$.}

    \label{fig:3}
\end{figure}

\textit{Target state preparation -} The analog-digital architecture allows us to deterministically prepare target states $\ket{\mathsf{T}}$, starting from the analog resource state $\ket{\mathsf{CS}}$. Since the underlying Hamiltonian of our system is of Fermi-Hubbard type, the four fermion Heisenberg chain provides a natural benchmark for the target state $\ket{\mathsf{T}}$. Nevertheless, our approach is not restricted to Hamiltonian eigenstates since digital configurability remains decoupled from the analog evolution.
To achieve the requisite all-to-all connectivity, we employ a minimal quantum circuit comprising a total of four digital gates: two inter-singlet and two intra-singlet gates on each of the isolated chained singlets. Each gate layer is separated by a reversal of the shuttle direction while the gate exponent $\alpha$ is kept constant for the full duration of the circuit (Figure \ref{fig:3}(a)). This ensures the spatial collimation of the target state within the lattice. 

We benchmark the protocol by comparing the measured correlations of the target state against theoretical predictions with no free parameters (Figure \ref{fig:3}(b)). The experimental results show very good agreement with theory, accurately tracking the evolution of the individual correlators as the gate exponent $\alpha$ is varied. A key feature is the emergence of ferromagnetic correlations of the $d=2$ correlator for $\alpha<1/4$.
To quantify the state preparation, we theoretically analyze the overlap of the generated target state with the ground state of the four-fermion Heisenberg chain $\ket{\mathsf{GS}}$. While the initial chained singlet state $\ket{\mathsf{CS}}$ approximates the ground state with an overlap of around $90\%$, the digital evolution can improve this value. As shown in Figure~\ref{fig:3}(c), applying the gate sequence with $\alpha=1/8$ increases the overlap to $>99\%$. This observation demonstrates that the analog-digital approach can efficiently approximate the ground state of few-body systems by digitally refining an analog resource state. Furthermore, this architecture provides a scalable pathway to larger systems, where increased connectivity can be achieved through additional layers of atom shuttling and collisional gates.

\begin{figure}
    \includegraphics[width=0.5\textwidth]{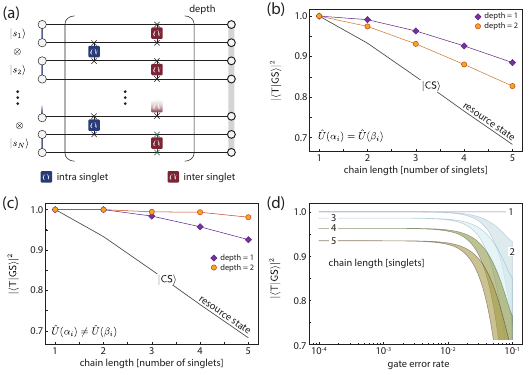}
         \caption{\textbf{Maximal ground state overlap of the target state for larger system size and deeper gate sequences.} (a) Circuit protocol for maximizing the ground state overlap of the output target state. The input state is a system with $N$ neighboring singlets which is manipulated by a sequence of alternating gates on the inter and intra singlet bonds. (b) The maximized ground state overlap of the initial state $\ket{\mathsf{R}_N}$ after manipulation with the circuit shown in (a). The gate exponent is optimized for maximal overlap, while being the same for all gates in the circuit. The baseline denotes the overlap of the initial state with the ground state without gate-based manipulation. (c) For arbitrary two-qubit partial \swap gate sequences the maximal ground state overlap increases for both circuit depths. (d) For gates with finite fidelity the maximal ground state overlap decreases with increasing gate error, while the preparation is robust up to error rates of $10^{-2}$. The shaded region indicates the cases where the combination of gate errors yields minimal or maximal overlap for all gates in a circuit.}
    \label{fig:4}
\end{figure}

\textit{Ground state overlap for increasing system size -} The analog-digital state preparation protocol is directly applicable to larger systems for deterministic state preparation. In this section, we theoretically investigate the capabilities of the introduced circuit protocol to prepare the approximate ground state of longer Heisenberg spin chains. The general brick-wall circuit structure applies two-qubit \swapa gates to neighboring atoms, alternating between inter-singlet bonds (connecting different singlets) and intra-singlet bonds (within initial singlets) as shown in the schematic in Figure \ref{fig:4}(a). The connectivity in larger systems is implemented with the use of deeper circuits that connect distant sites through the application of multiple gates between neighbouring atoms. This is essential for ground state preparation in the Heisenberg model, which is a global singlet of the system with characteristic long-range correlations \cite{qin_hubbard_2022}. To quantify the protocol, we theoretically investigate the ground state preparation fidelity in two scenarios: using a fixed partial \swap gate throughout the circuit (Figure \ref{fig:4}(b)) and allowing for time-dependent interaction strengths enabling arbitrary \swapa gates in each gate layer (Figure \ref{fig:4}(c)). We apply the quantum circuit to an initial state of $N$ neighbouring singlets ($\ket{\mathsf{CS}}_N=\bigotimes_{i=1}^N \singlet_i$). To find the optimal settings for the quantum circuit, we maximize the final state overlap with the anti-ferromagnetic ground state of the Heisenberg chain for circuits with two ($\text{depth}=1$) and four ($\text{depth}=2$) gate layers (Figure \ref{fig:4}(b),(c)). The results show that the analog-digital approach enhances the ground state overlap even in the case of longer chains. Especially with increasing system size, the gate based preparation significantly improves the ground state overlap with respect to the resource state $\ket{\mathsf{CS}}_N$.
Note that in the scenario of fixed gates the four gate layer circuit is outperformed by the two gate layer circuit, as the additional gate layers reverse part of the initially created long range correlations for a fixed gate exponent.  
To account for experimentally realistic scenarios of imperfect gate-operations, we investigate the fidelity of ground state preparation against error rates for the case of two gate layers with free gate exponents $\alpha$ (Figure \ref{fig:4}(d)). Under these experimental conditions, state preparation remains robust for gate errors up to $10^{-2}$, beyond which the overlap decreases rapidly. This demonstrates the feasibility of preparing the approximate ground state with digital methods for the Heisenberg chain in current Fermi-Hubbard quantum simulation platforms \cite{xu_neutral-atom_2025} with state-of-the-art two-qubit collisional gates \cite{zhang_scalable_2023,bojovic_high-fidelity_2026,kiefer_protected_2026}. While creating long-distance correlations in longer chains relies on the same protocols demonstrated for the four-particle system, the readout can be likewise extended to measure distant correlations.

\textit{Conclusion and Outlook -} We demonstrate the configuration of strongly correlated four-particle states with a combined analog-digital protocol in a fermionic optical lattice. Our approach uses isolated four-fermion instances within a large-scale optical lattice system as an analog resource state $\ket{\mathsf{CS}}$. The digital component is realised by combining deterministic atom rearrangement via topological pumping with tunable collisional gates to program and measure spin-correlations of a targeted final state $\ket{\mathsf{T}}$. To showcase the capabilities of the method, we initialise the ground state of the four-particle Heisenberg chain in a specific implementation of a quantum circuit.

The integration of analog and digital techniques establishes a versatile framework for the exploration of strongly correlated many-body states and quantum phases characterised by non-local correlations. Concatenating multiple layers of preparation steps could enable further refinement of highly entangled states and the measurement of higher-order correlations, essential for the microscopic identification of order parameters and phase transitions in systems that remain computationally inaccessible with classical numerics. The separation between bulk Hamiltonian evolution and gate-based digital refinement moreover enables the preparation of states beyond ground- or eigen-states of the Hamiltonian, including graph states \cite{hein_entanglement_2006} and symmetry-protected topological phases \cite{senthil_symmetry-protected_2015}. In contrast to the connected spin correlators accessed in quantum gas microscopes \cite{boll_spin-_2016,parsons_site-resolved_2016,mazurenko_cold-atom_2017}, the symmetry-selective measurement protocol gives access to the isotropic two-particle correlator, which encodes the orbital symmetry of the pair wavefunction and can be used to measure the global symmetry of the many-body state. This provides a complementary avenue for the detection of non-local order parameters, including topological order~\cite{pauw_detecting_2024} and superconducting pairing correlations encoded in off-diagonal long-range order \cite{mark_efficiently_2025,weitenberg_protocols_2026}.

In the context of quantum information science, our hybrid architecture addresses critical challenges in connectivity and scalability for fermionic systems \cite{bravyi_fermionic_2002,gonzalez-cuadra_fermionic_2023,calliari_programmable_2026,wei_universal_2026}. The deterministic control over atom positions and entanglement offers an efficient platform for exploring algorithmic problems \cite{cerezo_variational_2021} tailored to the symmetries of electronic systems ranging from condensed matter physics to quantum chemistry \cite{cao_quantum_2019}. Expanding these four-particle modules into larger, interconnected registers will facilitate the exploration of generalized error-correction schemes and the execution of complex circuits, addressing the gap between specialized quantum simulation and universal fermionic quantum computation.
\section*{Acknowledgments}
We acknowledge stimulating discussions with Lisa Peters, Giacomo Bisson, Samuel Jele, and Felix Borchers. We thank Christof Weitenberg and Jacob Fricke for critical reading of a previous version of the manuscript.We thank Alexander Frank for assistance with electronics equipment.
We acknowledge funding by the Swiss National Science Foundation (Grant No.~200020\_212168, Advanced grant TMAG-2\_209376, 20QT-1\_205584, as well as Holograph UeM019-5.1).
Y.K.~acknowledges funding from the ETH Postdoctoral Fellowship 24-2 FEL-035.

\bibliographystyle{unsrt}

\clearpage

\setcounter{figure}{0} 
\setcounter{equation}{0}

\renewcommand{\figurename}{Extended Data Fig.}
\renewcommand{\tablename}{Extended Data Table}
\renewcommand\thefigure{\arabic{figure}} 
\renewcommand\thetable{\arabic{table}} 
\renewcommand\theequation{M\arabic{equation}} 
\section*{Methods}

\renewcommand{\thesection}{}

\subsection{lattice potential and topological pump}
\label{SMsec:latPot}
\begin{figure}
    \includegraphics[width=0.5\textwidth]{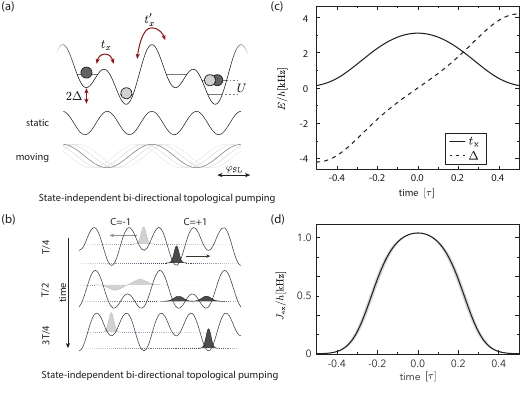}
         \caption{\textbf{Optical lattice parameters and topological pumping} (a) Dynamical optical lattice potential given by the superposition of a static short wavelength lattice and a moving long wavelength lattice. The optical potential is given by the inter- and intra dimer tunnelling $t_x$ and $t_x^\prime$ and the energy site offset $\Delta$, both tunable by adjusting the relative phase $\varphi_{\text{SL}}$ of the two lattices. (b) Periodic modulation of these parameters gives rise to topological pumping of localised Wannier orbitals of which the direction is given by the Chern number $C$ of the Bloch bands of the lattice. (c) Dynamical evolution of the interdimer tunneling $t_x$ and the energy site offset $\Delta$ accross one period of the topological pump $\tau$ for fixed lattice parameters $[V_\text{X},V_{\text{Xint}},V_\text{Y},V_\text{Z}]/\text{E}_{\text{rec}}=[10.0,1.1,31.0,30.0]$. (d) Exchange energy $J_{\text{ex}}$ during one half topological pump cycle for an Hubbard interaction $U=780$Hz. The grey shaded area denotes the exchange energy variation for tunnelling deviations of $\pm10\%$.}
    \label{SMfig:1}
\end{figure}

The optical lattice in our setup is generated by a single red-detuned laser with $\lambda=1064\,\text{nm}$, retro-reflected in all three spatial dimensions. Along the $x$-direction, a second beam ($X_\mathrm{int}$) is superimposed to create an interference pattern with the beam in the $z$-direction. The relative phase of the $X_\mathrm{int}$ and $Z$ lattice beams $\varphi_\text{SL}(\tau)$ can be adjusted dynamically, resulting in the time-dependent potential given by   
\begin{equation}
\begin{split}\label{eq:potential}
    V(&x,y,z,\tau) = \\
    &-V_\mathrm{X}\cos^2(kx+\theta/2)\\
    &-V_\mathrm{Xint}\cos^2(kx)\\
    &-V_\mathrm{Y}\cos^2(ky)\\
    &-V_\mathrm{Z}\cos^2(kz)\\
    &-\sqrt{V_\mathrm{Xint}V_\mathrm{Z}}\cos(kz)\cos(kx+\varphi_{\text{SL}}(\tau))\\
    &-I_\mathrm{XZ}\sqrt{V_\mathrm{Xint}V_\mathrm{Z}}\cos(kz)\cos(kx-\varphi_{\text{SL}}(\tau)),
\end{split}
\end{equation}
where $k=2\pi/\lambda$, the imbalance factor is $I_\text{XZ}$, and $\{V_\mathrm{X},V_{\mathrm{X}_\text{int}},V_\mathrm{Y},V_\mathrm{Z}\}$ denote the lattice depth of the individual beams with the values used in the experiment listed in Extended Data Table \ref{tab:lattice_depths}. The tunnelling dynamics along the $y$- and $z$-directions are frozen out on relevant timescales by choosing a sufficiently deep lattice potential in these spatial directions, effectively creating independent one-dimensional arrays along the $x$-direction. Each one-dimensional system is characterised by the tunable bias $\Delta$ and tunnellings $t_x$ and $t'_x$ (see Extended Data Fig.~\ref{SMfig:1}a). 
Cyclic modulation of $t_x$($t_x'$) and $\Delta$ is implemented by linearly ramping the superlattice phase $\varphi_\text{SL}(\tau)$ with an acousto-optic modulator acting on the $V_\mathrm{Xint}$ beam~\cite{walter_quantization_2023}.

{\renewcommand{\arraystretch}{1.2}
\begin{table*}[ht]
    \centering
    {\fontfamily{phv}\selectfont 
    \begin{tabular}{lcllll}
        \hline\hline
        & $V_\mathrm{X}$[$E_\mathrm{rec}$] & $V_\mathrm{Xint}$[$E_\mathrm{rec}$] & $V_\mathrm{Y}$[$E_\mathrm{rec}$] & $V_\mathrm{Z}$[$E_\mathrm{rec}$] & $I_\mathrm{XZ}$\\
        \hline
        Fig.~\ref{fig:2} measurement of resource and propagated singlet chains & 10.16(8)$^{\text{\tiny{a}}}$ & 1.25(3) & 29.07(5) & 27.08(7) & 0.804(1)\\
        Fig.~\ref{fig:3} programm four-fermion target state & 10.12(25) & 1.25(13) & 29.13(47) & 27.14(48) & 0.800(1)\\
        \hline
    \end{tabular}
    }
    \caption{\textbf{Lattice depths used in the experiment.} $^{\text{\tiny{a}}}$The errors in the brackets correspond to the statistical standard error.}
    \label{tab:lattice_depths}
\end{table*}}
The periodic modulation of the lattice potential can be considered as a static `short' lattice with a moving `long' lattice, describing a Thouless pump~\cite{citro_thouless_2023} (Extended Data Fig.~\ref{SMfig:1}(b)).
Compared to its classical counterpart, the quantum nature of the Thouless pump is manifested in its directional dependence of the motional states. In a band-structure picture, the pump in a bipartite one-dimensional lattice potential features two topologically distinct bands. The topological properties of these bands are characterised by Chern numbers. These numbers are derived by mapping the space- and time-periodic Hamiltonian onto a time-independent 2D Harper-Hofstadter-Hatsugai (HHH) model, which incorporates both a real and a synthetic dimension. The two lowest bands of the HHH model have Chern numbers $C = \pm 1$ giving rise to quantised transport of two lattice sites per period in opposite directions~\cite{oka_floquet_2019,walter_quantization_2023,zhu_splitting_2025}.

In contrast to bichromatic setups of topological pumps~\cite{nakajima_topological_2016,lohse_thouless_2016}, the depth of the `moving' lattice in our case is also periodically modulated. This modulation occurs automatically due to the time dependence of the two terms proportional to $\sqrt{V_\mathrm{Xint}V_\mathrm{Z}}$ in Eq.~\ref{eq:potential} and requires no change in laser intensities.
The modulation ensures a smoother time evolution of the topological bandgap and it increases the duration of the superexchange interaction (width of the feature in Extended Data Fig.~\ref{SMfig:1}d).

\subsection{Realisation of {\sc $($swap$)^\alpha$} gates in the low U regime}
\label{SMsec:swap}
The dynamical gate mechanism is implemented by adiabatically varying the tunneling $t$ and energy site offset $\Delta$ of a fermionic double-well. The operation proceeds by sweeping the potential from a negative to a positive staggered configuration, transitioning through a balanced double-well where $\Delta/t = 0$. In the singular limit of vanishing Hubbard interactions ($U = 0$), the Hamiltonian possesses chiral symmetry, which protects the adiabatic evolution of a dark state $|\Psi(\theta)\rangle = \cos(\theta/2)|s\rangle + \sin(\theta/2)|D_-\rangle$, with mixing angle $\theta$ defined as $\cot{(\theta/2)} = -\Delta/t$, residing exactly at zero energy. Under these non-interacting conditions, the gate is purely geometric; a complete evolution in the parameter space of the mixing angle $\theta\in[0,2\pi]$ results in a quantum holonomy, where the state acquires a geometric phase $\gamma = -\pi$ \cite{kiefer_protected_2026}.

The introduction of finite Hubbard interactions ($U \neq 0$) breaks the chiral symmetry and shifts the dark state from the zero-energy manifold. Consequently, the system acquires a dynamical phase $\delta = \int E_\psi(\tau) d\tau$ in addition to the geometric contribution. In the direct exchange regime ($|U| \le t$), the energy of the evolving state is determined by the interaction strength in the first order, ensuring that the accumulated dynamical phase is directly proportional to $U$. This scaling contrasts fundamentally with the superexchange regime ($U \gg t$), where the effective coupling $J \approx 4t^2/U$ arises from second-order perturbative processes. While superexchange operations are highly sensitive to fluctuations in the tunneling amplitude $t$, the dynamical gate leverages the first-order dependence on $U$ to provide a robust phase acquisition. This resilience is further reinforced by fermionic exchange anti-symmetry, which ensures the triplet manifold $\mathcal{T}$ remains at zero energy and serves as a stable phase reference. The total relative phase $\varphi = \gamma + \delta$ facilitates high-fidelity entangling operations and SWAP gates.

The Hamiltonian governing the evolution in the $\{\mathcal{T, S}\}$ basis is defined as:
\begin{equation}
\hat{H}_{\mathcal{T, S}}(\tau) = \begin{pmatrix}
\mathbf{0}_3 & \mathbf{0}_3 \\
\mathbf{0}_3 & \begin{matrix} U & 2\Delta(\tau) & -2t(\tau) \\ 2\Delta(\tau) & U & 0 \\ -2t(\tau) & 0 & 0 \end{matrix}
\end{pmatrix}
\end{equation}
\subsection{Singlet purification}
\label{SMsec:singletDist}
\begin{figure}
    \includegraphics[width=0.5\textwidth]{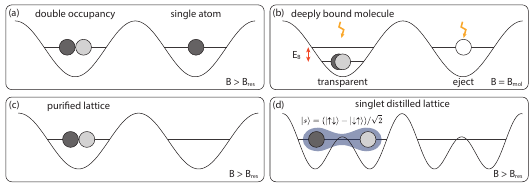}
         \caption{\textbf{Singlet purification scheme} (a-d) Experimental sequence used to perform singlet purification in the three dimensional optical lattice using the formation of deeply bound molecular states that remain transparent to resonant light. }
    \label{SMfig:3}
\end{figure}
One of the central advancements of our scheme is the ability to isolate instances in which exactly two atoms of opposite spin occupy the double-well potential. Our experimental sequence begins with the preparation of a deeply degenerate Fermi gas of $^{40}$K at $T=0.10(2)T_F$ in a crossed optical dipole trap, using a balanced mixture of $\ket{F=9/2, m_F=-9/2}$ and $\ket{F=9/2, m_F=-7/2}$ Zeeman substates. To favor the formation of doubly occupied (DO) sites in the 'long' lattice, we load the atoms into the lowest band of an optical three-dimensional checkerboard lattice while maintaining attractive Hubbard interactions. This configuration enhances the probability of DOs. However, due to finite temperature of the bulk system, a significant fraction of lattice sites ($\sim 40\%$) remains single occupied (Figure \ref{SMfig:3}(a)). To achieve the required high-fidelity initialization, we therefore perform singlet purification to remove single-atom impurities. Our strategy exploits selective removal of single atoms without disturbing the DO lattice sites. The process is implemented via the following steps. First we perform magnetoassocation of the DO lattice sites. For this, we adiabatically ramp the magnetic field through a broad Feshbach resonance located at $B_{res}=202.1$G to a final field of $B_{mol}=10$G in 5ms. At $B_{mol}$, the molecules occupy a deeply bound molecular two-body state which is inherently protected from three-body loss (Figure \ref{SMfig:3}(b)). After a 30ms hold time, we apply a $\SI{140}{\us}$ pulse of resonant light tuned to the the middle of the resonances of $\ket{m_F=-9/2}$ and $\ket{m_F=-7/2}$. This pulse clears unpaired atoms from the singly occupied lattice sites, while the molecular pairs remain transparent to the light pulse. Subsequently, the magnetic field is ramped back above the Feshbach resonance to restore the atomic two-body states (Figure \ref{SMfig:3}(c)). The next step adiabatically splits the DOs into atomic singlets. For this we transition from the single-site checkerboard lattice to a double-well scenario by simultaneously ramping up a short wavelength lattice and ramping down the long wavelength lattice. This process adiabatically connects the DO state to the local ground state of the double-well. If the interactions are chosen repulsive ($U=\SI{2000}{Hz}$) during the splitting process, atomic singlets $\ket{s}_{ij}$ are obtained on all lattice sites previously occupied by two atoms (Figure \ref{SMfig:3}(d)). Upon completion of the purification sequence, we obtain a many-body system consisting of approximately 25000 atoms with a singlet purity exceeding $80\%$ (see Methods \ref{SMsec:signalIsolation}). This preparation provides the necessary starting point for subsequent digital manipulations and spin-correlator engineering. 
\subsection{Initial state characterisation}
\label{SMsec:initialState}
\subsubsection{global singlet fraction}
\begin{figure}
    \includegraphics[width=0.5\textwidth]{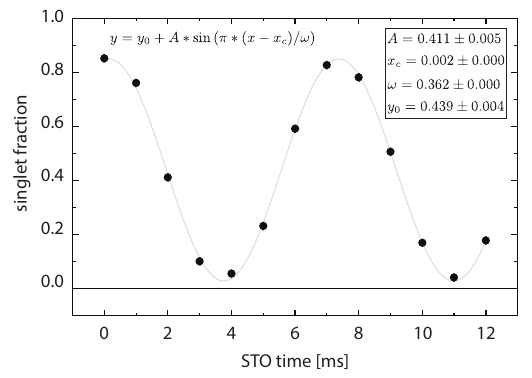}
         \caption{\textbf{Global singlet fraction measurement} Following the singlon cleaning we immediately apply a STO to extract the singlet fraction. The fraction is obtained by taking the amplitude of the sinusodial fit of which the model is written in the top part of the plot. The fraction is $2*A = 0.822(1)$. The error bars denote the standard error for i=16-20 repititions. }
    \label{SMfig:2}
\end{figure}

We characterise the initial state after the removal of isolated atoms with two different experiments. The global singlet fraction is obtained by inducing coherent singlet-triplet oscillations immediately after the singlon cleaning. The data is fit with a sine function of which the two times the amplitude denotes the total singlet fraction. The total singlet fraction is $2*A = 0.822(1)$ as shown in Figure \ref{SMfig:2}
\subsubsection{Four-particle chains}
\begin{figure}
    \includegraphics[width=0.5\textwidth]{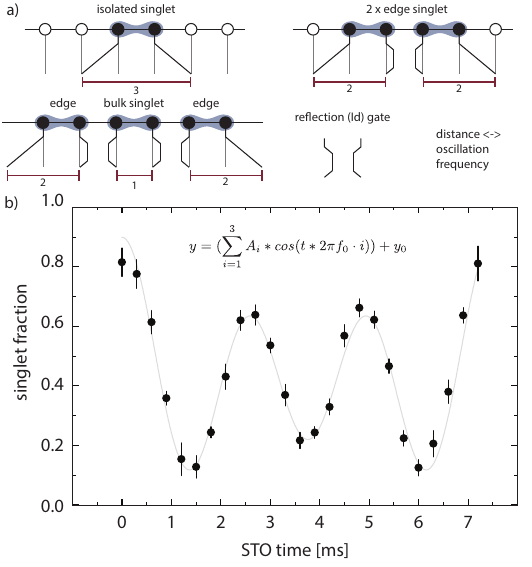}
         \caption{\textbf{Four-particle chain characterisation} (a) schematic illustration of the circuit applied to the system to obtain the fraction of the four particle chains.  A lower bound for the number of four particle chains can be given by taking twice the amplutde . The error bars denote the standard error for i=16-20 repititions. }
    \label{SMfig:2_2}
\end{figure}

The fraction of four particle chains is estimated with a quantum circuit composed of one shuttling cycle while applying an identity gate (see Figure \ref{SMfig:2_2}(a)). The gate is based on atomic collisions and therefore realises a process conditioned on the presence of another atom. Depending on the filling of the chain three different scenarios can happen. (i) Isolated singlets are surrounded by vacancies on both sides are separated by three lattice sites after the circuit. (ii) Edge singlets that are surrounded by only one neighbour and a vacancy attribute oscillations with a frequency $f=2f_0$ corresponding to distance $d=2$. (iii) Bulk singlets are surrounded by another singlet on both sides will attribute oscillations with frequency $f_0$. Depending on the fitted amplitudes of the oscillations at these frequencies we can determine an upper and a lower bound to the fraction of four particle chains. 
The fitted populations are $2*A_1 = 0.12(2)$ (bulk singlets), $2*A_2= 0.24(2)$ (edge singlets) and $A_3=0.56(2)$ (isolated). 

From this, we infer a lower bound of $\ket{\mathsf{CS}}_{\text{frac}}^{\text{lower}}=0.12(2)$ under the assumption that all of the bulk singlets are surrounded by two edge singlets. This assumption is only correct, if we assume that the maximum chain length is six sites (3 singlets). It is very likely that longer chains are present in the experiment which would lead to a larger fraction of four-particle chains. The upper bound is given under the assumption that all bulk singlets are located in a single chain. In this case one could directly substract just two edge singlets from the $d=2$ amplitude essentially directly corresponding to an upper bound of $\ket{\mathsf{CS}}_{\text{frac}}^{\text{upper}}=0.24(2)$. The STO contrast shown in Figure \ref{fig:2} is consistent with these bounds. We measure an amplitude of $A_{\text{Fig2}}=0.065(3)$ indicating that most of the bulk singlets are surrounded by two neighbours.

\subsection{Extracting spin-correlations from four-fermion chains}
\label{SMsec:signalIsolation}
Our system consists of a two-dimensional array of isolated one-dimensional tubes with superlattice structure. Each experimental data point averages over this array, where individual tubes have varying particle occupancies due to the inhomogeneous density distribution of the atomic cloud. The tubes at the edges of the system have lower occupancy than those near the center. The fraction of isolated two-particle systems (neighboring double-wells unoccupied) and four-particle chains (one neighboring double-well occupied, the other unoccupied) remains stable across experimental cycles. At the same time, singlet purification removes isolated single atoms (Methods \ref{SMsec:singletDist}), while chains with $N\geq6$ make up less than $5\%$ of the system. 

The two-particle correlator measurements are divided into two classes: those where isolated singlets (two-particle chains) contribute to the local readout signal ($\langle C_{12}\rangle,\langle C_{34}\rangle,\langle C_{13}\rangle,\langle C_{24}\rangle$), and those where they do not ($\langle C_{23}\rangle,\langle C_{14}\rangle$). In the latter case, the obtained signal is purely from the four-particle chains, while in the first case we need to subtract the background correlation contributed by the isolated singlets. 

The programmable quantum circuit encodes globally synchronized atom shuttling, where only the presence of two atoms on a double-well activates changes in internal degrees of freedom through atomic collisions. Consequently, all Bell pairs experience identical displacements, and only their embedding within a larger chain modifies the system's entanglement structure. The circuits are designed such that only one auxiliary site is needed on each end of a chain to program the two particle correlations. Singlet purification is essential here because it creates unoccupied double-wells adjacent to two and four particle chains. All isolated systems are separated by at least one unoccupied double-well, providing two auxiliary sites for neighboring systems. Therefore, the two and four particle systems do not interact during the circuit protocol.
When we measure the two particle correlators of the four particle chains, where also isolated singlets contribute to the singlet-triplet oscillation signal, the corrected signal is obtained from the singlet-triplet oscillation difference between a reference measurement (singlet-triplet oscillation of all initial singlet pairs) and the programmed four particle circuit. 

As a consequence of individual chains being isolated from each other, the Hilbert space of the total system factorizes and can be written as
\begin{equation}
    \mathcal{H} = \bigotimes_k \mathcal{H}_{C_k}, 
\end{equation}
where the disjoint subspaces $\mathcal{H}_{C_k}$ are two and four particle chains, as well as empty sites. The initial state preparation reduces the basis that span $\mathcal{H}_{C_k}$, which can be written as 
\begin{align*}
    \mathcal{H}_{C_k}^{0} = \{&\ket{0,0}\},\\ 
    \mathcal{H}_{C_k}^{2} = \{&\ket{\uparrow,\downarrow},\ket{\downarrow,\uparrow}\}, \\
    \mathcal{H}_{C_k}^{4} = \{&\ket{\uparrow\uparrow\downarrow\downarrow},\ket{\uparrow\downarrow\uparrow\downarrow},\ket{\uparrow\downarrow\downarrow\uparrow}, \\ &\ket{\downarrow\uparrow\uparrow\downarrow},\ket{\downarrow\uparrow\downarrow\uparrow},\ket{\downarrow\downarrow\uparrow\uparrow}\}. 
\end{align*}
Total initial states of the system are thus expressed as 
\begin{align*}
    \ket{\psi} = \bigotimes_{k,n} \ket{\psi^n_k} 
\end{align*}
where the $\ket{\psi_k^n}\in \mathcal{H}_{C_k}^n$. Based on the circuit protocol where different subspaces are independent, the global unitary acting on the total system also factorizes 
\begin{align*}
    U = \bigotimes_k U_k
\end{align*}
with each $U_k$ acting only in $\mathcal{H}_{C_k}^n$. Similarly in the Hamiltonian formalism the global Hamiltonian is given by the sum of all local Hamiltonians $H_k(t)$ where for all $k=k'$ the commutator is $[H_k(t),H_{k'}(t')]=0$. As a consequence, the total density matrix of the system $\rho$ can be expressed as 
\begin{align*}
    \rho=\bigotimes_k U_k\rho_k U_k^\dagger,
\end{align*}
and the individual subsystems are independent and their signals separable.

\end{document}